\documentclass[prd,aps,twocolumn,preprintnumbers,superscriptaddress,showpacs,floatfix,nofootinbib]{revtex4-1}
\pdfoutput=1

\usepackage{tikz}
\usepackage{booktabs}
\usepackage[toc,page]{appendix}
\usepackage{amsmath,bbm,array,amsfonts,float,mathtools,amsthm}
\usepackage{multirow}
\usepackage{microtype}
\usepackage{hyperref}
\usepackage[toc,page]{appendix}
\usepackage[utf8]{inputenc}
\usepackage[autostyle=true]{csquotes}
\usepackage{todonotes}
\usepackage{graphicx}
\usepackage{environ}
\usepackage[framemethod=Tikz]{mdframed}

\usetikzlibrary{decorations.pathmorphing}
\usetikzlibrary{decorations.pathreplacing}
\usetikzlibrary{decorations.markings}
\usetikzlibrary{shapes, shapes.geometric, shapes.symbols, shapes.arrows, shapes.multipart, shapes.callouts, shapes.misc}
\usetikzlibrary{shadows,fit,trees,calc,patterns,arrows}
\tikzset{snake it/.style={decorate, decoration=snake, segment length=2mm}}
\tikzset{7brane/.style={circle, draw=black, fill=black,ultra thick,inner sep=1.5 pt, minimum size=1 pt,}, c/.default={4pt}}
\tikzset{cross/.style={cross out, draw=black,ultra thick, minimum size=2*(#1-\pgflinewidth), inner sep=0pt, outer sep=0pt}, cross/.default={5pt}}

\numberwithin{equation}{section}
\numberwithin{figure}{section}
\numberwithin{table}{section}
\hypersetup{
    colorlinks,
    linkcolor={red!50!black},
    citecolor={blue!50!black},
    urlcolor={blue!80!black}
}

\begin{document}

\title{Quivers, Lattice Gauge Theories and Fractons}

\author{Sebasti\'an Franco}
\email{sfranco@ccny.cuny.edu}
\affiliation{Physics Department, The City College of the CUNY 160 Convent Avenue, New York, NY 10031, USA}
\affiliation{Physics Program and Initiative for the Theoretical Sciences, The Graduate School and University Center, The City University of New York 365 Fifth Avenue, New York NY 10016, USA}

\author{Diego Rodr\'iguez-G\'omez}
\email{d.rodriguez.gomez@uniovi.es}
\affiliation{Department of Physics, Universidad de Oviedo, C/ Federico Garc\'{i}a Lorca 18, 33007, Oviedo, Spain}
\affiliation{Instituto Universitario de Ciencias y Tecnolog\'{i}as Espaciales de Asturias (ICTEA), C/ de la Independencia 13, 33004 Oviedo, Spain}


\begin{abstract}

We argue that quiver gauge theories with $SU(N)$ gauge groups give rise to lattice gauge theories with matter possessing fractonic properties, where the lattice is the quiver itself. This idea extends a recent proposal by Razamat. This class of theories exhibit a $\mathbb{Z}_N$ 1-form global symmetry that can be used to classify their phases. The order parameter of this transition is the expectation value of Wilson loops, which correspond to mesonic operators in the underlying quiver gauge theory. We discuss how this perspective naturally fits with the deconstruction of a higher dimensional theory.

\end{abstract}

\maketitle

\section{Introduction}
\label{sec:intro}

Lattice models are ubiquitous in multiple branches of physics. Sometimes, they describe the actual structure of a system, e.g. in the case of the microscopic description of atoms in a solid. In other cases, they provide useful models and powerful computational tools, such as when describing the propagation of waves in a continuous medium as the long wave limit of a system of coupled harmonic oscillators. 

Lattice theories giving rise to fractonic phases of matter have generated considerable interest in the condensed matter community (see \cite{Pretko:2020cko} for review and references). These systems exhibit various interesting properties, related to the existence of sub-system symmetries, including excitations with restricted mobility and large number of vacua.

The insightful paper \cite{Razamat:2021jkx} identified a deep similarity between a class of quiver gauge theories and the lattice theories underlying fracton models, and explained how the aforementioned features arise in this context. Quiver gauge theories are quantum field theories whose gauge symmetry, matter content and interactions can be efficiently encoded in an oriented graph – called the {\it quiver}. Perhaps in another sign of the universality of the underlying ideas, precisely the type of quivers in \cite{Razamat:2021jkx} has been extensively studied in a seemingly distant corner of physics: gauge theories engineered on D-branes probing singularities \cite{Franco:2005rj}. For this reason, they have also naturally appeared in \cite{Geng:2021cmq}, where various D-brane realizations of fractons were introduced.

This type of setups can be understood within the general paradigm of deconstruction \cite{Arkani-Hamed:2001kyx,Arkani-Hamed:2001nha}. There, quivers describing $d+1$-dimensional gauge theories and with a $d’$ internal structure effectively lead, at low energies, to the emergence of $d’$ new dimensions, whose geometry is represented by the quiver. The quiver can be regarded as a lattization of the new dimensions. Deconstruction is a rather general framework, which can even lead to ``extra dimensions” without an obvious geometric interpretation (think e.g. about an 8 or X-shaped quiver dimension). Deconstruction therefore takes the notion of lattice theories to a higher level of abstraction. Instead of the lattice occupying discrete sites in some real dimensions, it lives in theory space and leads to the effective materialization of new dimensions.

In this note we will study a particular class of 4d quiver gauge theories encoded in 2d quivers with $SU(N)$ gauge groups. These theories were used in \cite{Arkani-Hamed:2001wsh} to deconstruct the 6d (2,0) and little string theories \cite{Seiberg:1997zk,Aharony:1999ks}. We will argue that the quiver itself defines a 2d lattice $U(1)$ gauge theory with charge $N$ matter given by the fractons described in \cite{Razamat:2021jkx}. Therefore, besides the standard 0-form global symmetries, the 2d lattice gauge theory has a 1-form $\mathbb{Z}_N$ global symmetry \cite{Gaiotto:2014kfa} under which Wilson lines (in the 2d lattice sense) are charged. This 1-form global symmetry characterizes the different phases of the 2d gauge theory. In the confined phase, which in 4d language corresponds to the baryonic branch and is expected to deconstruct the 6d theory \cite{Arkani-Hamed:2001wsh}, the 1-form symmetry remains unbroken and can be combined with the standard $\mathbb{Z}_N$ 1-form symmetry of the 4d gauge theory to give rise to the 1-form global symmetry of the 6d theory \cite{Bhardwaj:2020phs}. In the deconfined phase, which in 4d language corresponds to the mesonic branch, Wilson loops acquire a non-zero vacuum expectation value (VEV) and the 2d $\mathbb{Z}_N$ 1-form symmetry is broken.

\section{A lattice gauge theory Model for the chiral ring}

Quivers are oriented graphs in which nodes, edges and oriented loops correspond to gauge groups, matter fields and interactions between matter fields in a $d+1$-dimensional (supersymmetric) gauge theory, respectively. 

As in \cite{Razamat:2021jkx}, we  will focus on the class of 4d $\mathcal{N}=1$ supersymmetric gauge theories defined by the triangular quivers schematically shown in fig.\ref{Figure:TheoryInfo}. Every node corresponds to a gauge group and every arrow corresponds to a bifundamental chiral superfield. Furthermore, we take all gauge groups to be $SU(N)$. Every bifundamental field should therefore be regarded as an $N\times N$ matrix.

The quiver lives on a 2-dimensional torus, \textit{i.e.} it is periodically identified along two directions, whose lengths are respectively $L_1$ and $L_2$ nodes. In the general language of the introduction, these theories have $d=3$ and $d'=2$. This theory describes the low energy dynamics of a stack of $N$ D3-branes in Type IIB String Theory probing a $ \mathbb{C}^3/\mathbb{Z}_{L_1}\times \mathbb{Z}_{L_2}$ singularity. While not important for the current discussion, different actions of the orbifold group translate into different periodic identifications (see. e.g. \cite{Davey:2010px,Hanany:2010cx}). In the spirit of deconstruction, the ``quiver space” can be regarded as a lattized 2-dimensional space.

\vspace{-0.4cm}
\begin{figure}[h!]
\centering
\includegraphics[scale=.6]{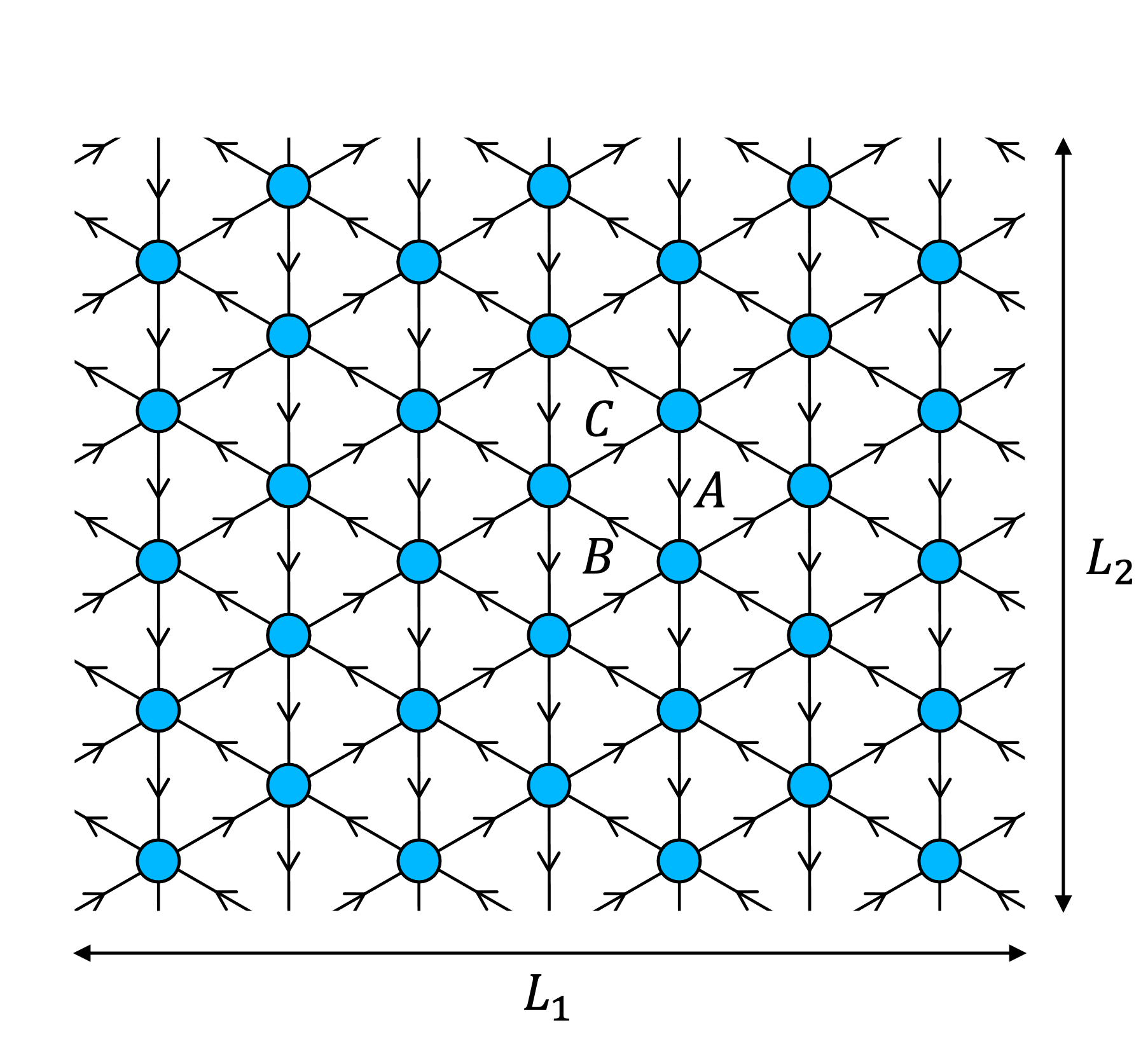}
\caption{Class of quiver theories under consideration. Every node is an $SU(N)$ gauge group. Arrows represent chiral superfields in the bifundamental representation of the nodes they connect.}
\label{Figure:Quiver}
\end{figure}

The 4d gauge theory also includes a superpotential, which is encoded in the periodic quiver and takes the form
\begin{equation}
\label{W}
    W=\sum_{\{i,j,k\}\in P}\lambda_{ijk} \,A_i\,B_j\,C_k\,,
\end{equation}
where $i$, $j$ and $k$ label the bifundamental fields and the sum is over the set $P$ of triangular plaquettes. The $\lambda_{ijk}$ are coupling constants.

\subsection{A $U(1)$ gauge symmetry on the torus}

Perturbatively, the global symmetry of the 4d gauge theory includes a $U(1)^{L_1\times L_2}$ group arising from $U(1)$ rotations at each node. Each of these $U(1)$ factors acts ``locally” at a node $i$ of the quiver, transforming incoming arrows by $e^{i\theta_i}$ and outgoing arrows by $e^{-i\theta_i}$. Thus, from the point of view of the quiver as a lattice, we have a 2d lattice gauge theory with gauge group $U(1)$.

However, anomalies break most of these $L_1\times L_2$ symmetries, so that only $L_1+L_2+{\rm gcd}(L_1,L_2)-1$ of these remain true quantum symmetries.\footnote{The $-1$ is due to the fact that the diagonal combination of these $U(1)$ groups decouples, since everything is neutral under it in a theory constructed in terms of bifundamental fields. A complete analysis of the global symmetries of these theories is presented in \cite{Razamat:2021jkx}.} A useful perspective on these theories is obtained by considering their String Theory embedding \cite{Douglas:1996sw,Morrison:1998cs}. In this UV completion, all the gauge groups are $U(N)$ at high energies and the theory includes compensating couplings which render all $U(1)$ factors non-anomalous \cite{Ibanez:1998qp}. These couplings give rise to a generalized Green-Schwarz mechanism, which leads to the spontaneously breaking of most of the $U(1)$'s, leaving behind precisely the expected $L_1+L_2+{\rm gcd}(L_1,L_2)-1$ global symmetries of the IR theory. Regardless of their fate, spontaneously broken or not, from this high-energy perspective there is indeed a $U(1)$ per site. Since these $U(1)$'s are IR free, we will think of them as global symmetries. We conclude that the theory in fig.\ref{Figure:TheoryInfo}, understood in the broad sense above including the compensating couplings, can be interpreted as a latticized/deconstructed $U(1)$ gauge theory on the torus.

\begin{figure}[h!]
\centering
\includegraphics[width=0.9\linewidth]{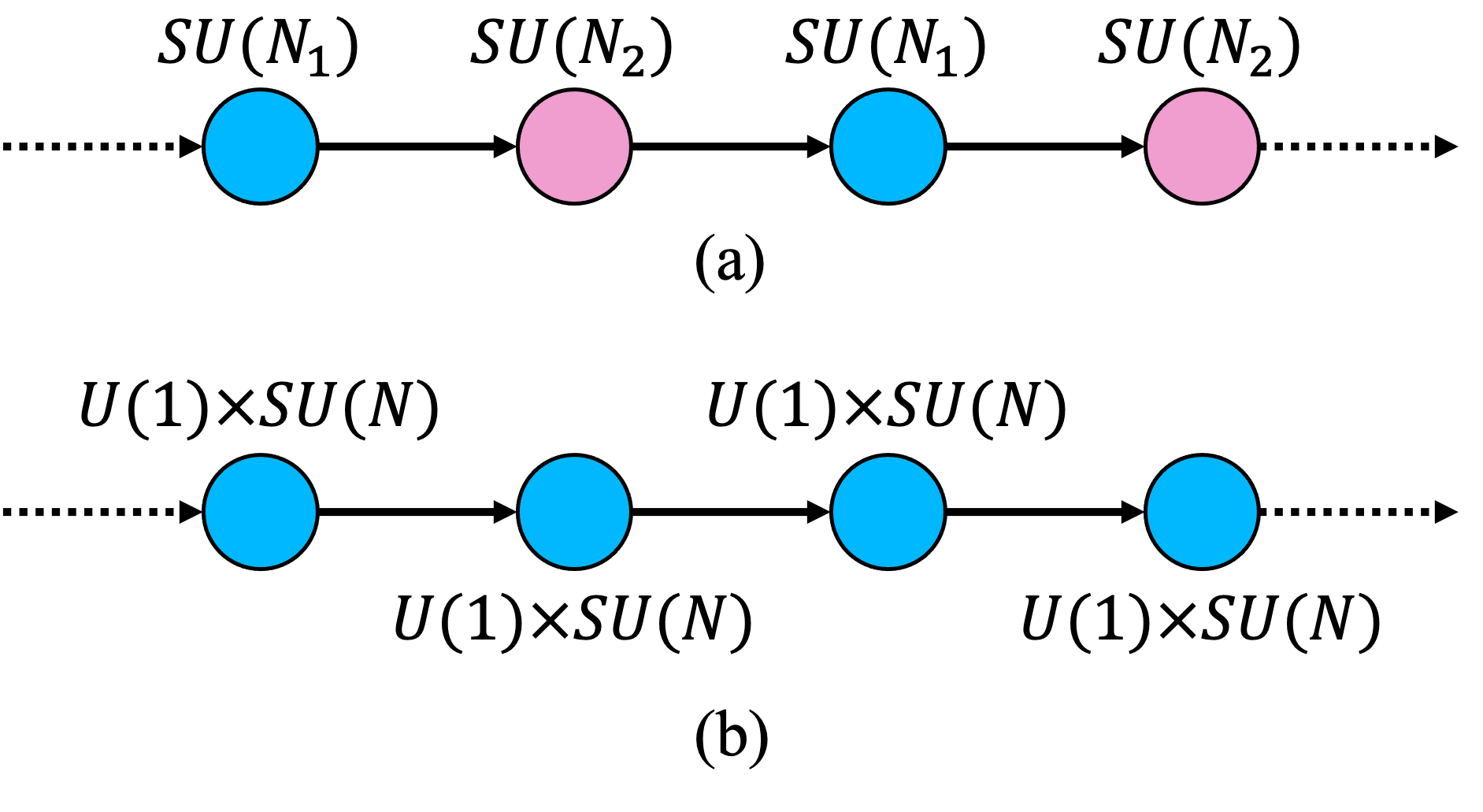}
\caption{a) Basic quiver producing an $SU(N_2)$ gauge theory in a deconstructed dimension. b) Quiver leading to a $U(1)$ gauge theory in a deconstructed dimension.}
\label{Figure:deconstruction}
\end{figure}

It is interesting to compare this example with the most familiar incarnations of deconstruction \cite{Arkani-Hamed:2001kyx,Arkani-Hamed:2001nha}. These involve quivers, which for simplicity of the presentation we depict as 1-dimensional in fig.\ref{Figure:deconstruction}.a, with a series of alternating $SU(N_1)$ and $SU(N_2)$ gauge groups. Assuming that the dynamical scales of the two types of gauge group are such that $\Lambda_1 \gg \Lambda_2$, the result at low energies is an $SU(N_2)$ gauge theory on the emergent dimension generated by the $SU(N_1)$ dynamics. The structure of theories we are considering is actually very similar. As illustrated in fig.\ref{Figure:deconstruction}.b for the case of a single deconstructed dimension, we can think that there are two types of groups, $SU(N)$ and $U(1)$, and we are left with a $U(1)$ gauge theory in the dimension generated by the $SU(N)$ dynamics. In this case, instead, the two groups are associated to the same node of the quiver. As previously mentioned, this fits with the D-brane realization of these gauge theories, where the gauge group associated to every node in the quiver is $U(N)=U(1)\times SU(N)$, with the $U(1)$ factor flowing to zero coupling in the IR (and in addition, from this perspective, the majority of them spontaneously broken as described above). Note that quivers including the $U(1)$ factors, \textit{i.e.} based on $U$ rather than $SU$ groups, have been shown to be relevant in other instances related to deconstruction (see \textit{e.g.} \cite{Hayling:2017cva,Hayling:2018fmv})

\subsection{Local excitations}

From a 4d perspective there are two types of gauge invariant (under the 4d $SU(N)$ gauge symmetries) operators:  mesons -- traces of concatenated bifundamentals-- and baryons -- determinants of bifundamentals.\footnote{We restrict to the simplest baryons. It is possible to consider more general baryonic operators, given by determinants of strings of concatenated bifundamentals. From a 2d viewpoint, they are analogous to charged operators inserted at different points, connected by a Wilson line to make the configuration gauge invariant.}

From the 2d point of view, baryons are associated to links of the lattice, and can therefore be interpreted as local excitations \cite{Razamat:2021jkx}. While baryons are gauge invariant from the 4d point of view, they have charge $N$ under the 2d $U(1)$ gauge symmetry. 

Since we will be particularly interested in the 2d point of view, from now on we will refer to the baryons as the (charged) matter and concentrate, unless otherwise stated, on their 2d properties.

\subsection{Global symmetries}

The 2d lattice gauge theory has global (0-form) symmetries  which combine into a global symmetry group 
\begin{equation}
G_0=U(1)_R\times\frac{U(1)^{L_1}\times U(1)^{L_2}\times U(1)^{{\rm gcd}(L_1,L_2)}}{U(1)}\,.
\end{equation}
The non-$R$ factors of these symmetries act as subsystem symmetries and were used in \cite{Razamat:2021jkx} to argue that matter fields exhibit fractonic properties. In particular, individual baryons cannot propagate on the lattice, while suitable combinations can move in certain directions.

On general grounds, there is a 2d 1-form global symmetry associated to the 2d gauge symmetry \cite{Gaiotto:2014kfa}. Given that there is matter with charge $N$ under the gauge symmetry, the 1-form global symmetry group is $G_1=\mathbb{Z}_N$.

It is interesting to note that, from the point of view of the low energy gauge theory, where one would simply say that most $U(1)$'s are anomalous, the dangerous mixed anomaly is proportional to $N$, and so a $\mathbb{Z}_N$ subgroup remains unbroken. This $\mathbb{Z}_N$ can actually be regarded as the center of the $SU(N)^{L_1\times L_2}$ gauge symmetry (and therefore can be upgraded to a gauge symmetry in 4d as well). Thus, we may equally phrase the discussion in the pure IR gauge theory, where the quiver gives rise to a 2d lattice $\mathbb{Z}_N$ gauge theory which carries the associated  $\mathbb{Z}_N$ 1-form global symmetry. Therefore, we can alternatively think that the $\mathbb{Z}_N$ 1-form global symmetry arises from the $\mathbb{Z}_N$ 2d lattice gauge theory given by center transformations at definite sites of the quiver.

\subsection{Wilson loops}

Mesons translate into oriented closed loops in the quiver, that is, closed paths in the lattice. From a 2d viewpoint, we interpret them as the Wilson loops associated to the $U(1)$ lattice gauge theory. Note that, as it should, they are neutral under the 2d $U(1)$ gauge symmetry. As for the global symmetries, Wilson loops are neutral under all the 0-form symmetries $G_0$. On the other hand, they are charged under the 1-form global symmetry $G_1$.

\begin{figure}[h!]
\centering
\vspace{-0.5cm}\includegraphics[scale=.6]{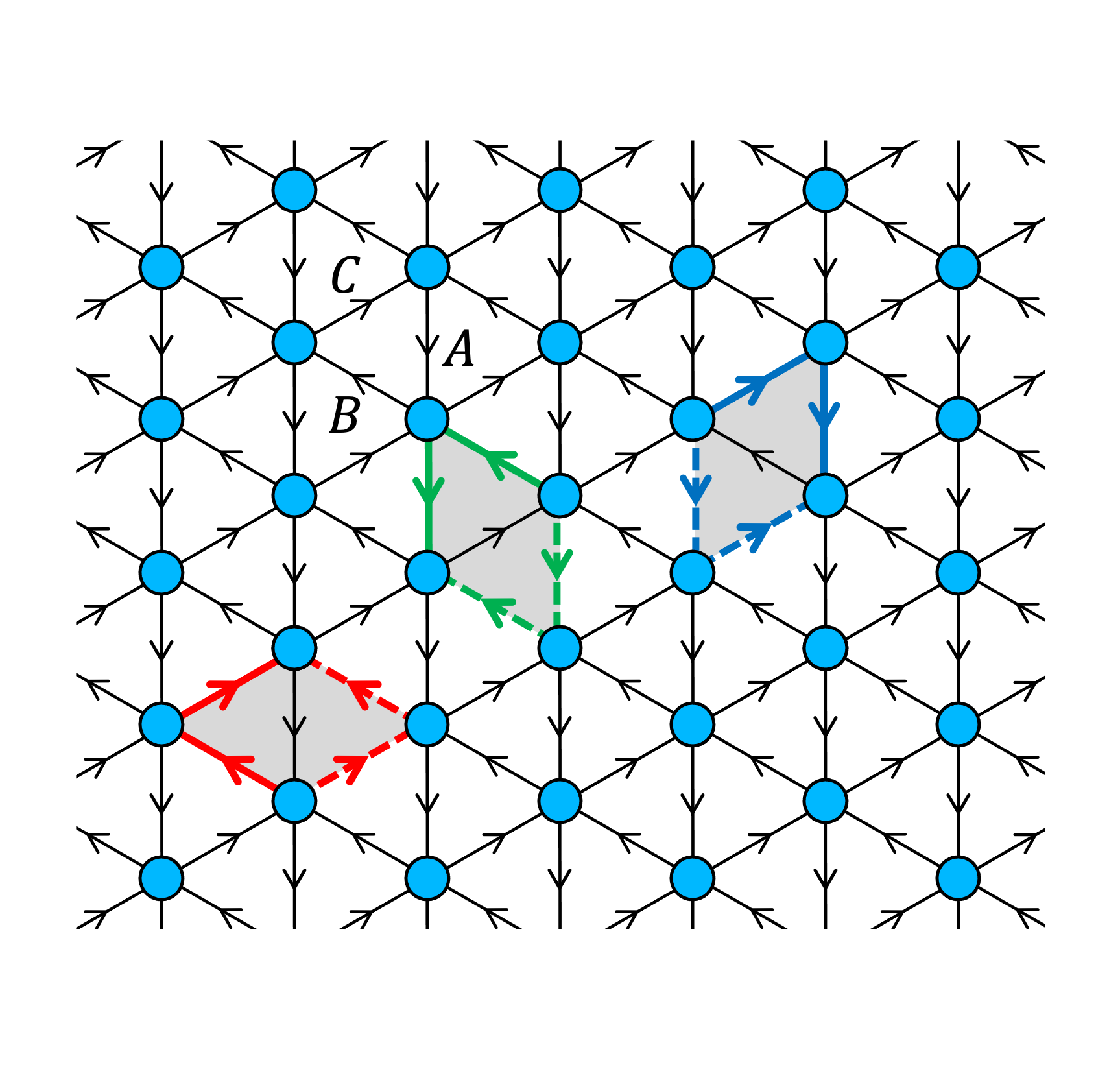}
\vspace{-0.6cm}\caption{Quiver and $F$-term relations: the solid red/green/blue paths are equivalent to the dashed red/green/blue paths.}
\label{Figure:TheoryInfo}
\end{figure}

In the chiral ring, the $F$-terms coming from the superpotential \eqref{W} must vanish. Note that in the class of theories we consider, which are defined by quivers on a torus, every field appears in exactly two adjacent plaquettes. Thus, the $F$-term equations take the general form 
\begin{equation}
B_p\,C_p=B_{p'}\,C_{p'}\,,\quad C_p\,A_p=C_{p'}\,A_{p'}\,,\quad A_p\,B_p=A_{p'}\,B_{p'}\,,
\label{F-term_relations}
\end{equation}
where we have introduced a new notation for the subindices that emphasizes that the two sides of each equation correspond to the two plaquettes that overlap at an $A$, $B$ or $C$ type field, respectively.\footnote{In this notation, the subindices indicate the plaquette that bifundamental fields belong to. While this notation is slightly less precise than the one used in equation \ref{W}, since the numbers of plaquettes and fields of each type are different, we hope it is self-explanatory and does not lead to confusion.} There is one such equation for every bifundamental field, and $p$ and $p'$ denote the two plaquettes that contain it. As commonly done in the literature, we have set the coupling constants $\lambda_{ijk}$ to be $+1$ for clockwise plaquettes and $-1$ for counterclockwise plaquettes. The relations in \ref{F-term_relations} imply the equivalence between open paths on adjacent plaquettes as illustrated in fig.\ref{Figure:TheoryInfo}. Repeated application of these relations leads to the equivalence of more general pairs of open paths sharing the same endpoints. In particular, they allow the Wilson loops to freely move along the 2d lattice.

The chiral ring relations have yet one more interesting consequence. In the chiral ring, there is a relation between the $N$-th product of a meson and the product of baryons of the constituents. For example, consider the simplest meson associated to a plaquette, $\mathcal{M}_p={\rm Tr}(A_p\,B_p\,C_p)$. Roughly speaking, there is a relation of the form $\mathcal{M}_p^N\sim {\rm det}A_p\,{\rm det}B_p\,{\rm det}{C_p}$. We interpret this relation as encoding the fact that $N$ Wilson lines are endable (in the language of \textit{e.g.} \cite{Heidenreich:2021xpr}) and can end on the charge $N$ baryons. This reflects the fact that the 1-form global symmetry is $\mathbb{Z}_N$.\footnote{The chiral ring relation between mesons and baryons is most naturally understood as endable Wilson lines from the point of view of the 2d $U(1)$ lattice theory. Had we taken the ``IR point of view" leading to a 2d $\mathbb{Z}_N$ lattice gauge theory, the baryons would be neutral under the gauge symmetry and the relation $\mathcal{M}_p^N\sim {\rm det}A_p\,{\rm det}B_p\,{\rm det}{C_p}$ would simply say that $N$ Wilson lines carry no charge and can annihilate into neutral matter.}

\subsection{Phases of the theory}

The 2d $\mathbb{Z}_N$ 1-form global symmetry characterizes the different phases of the 2d lattice theory. The order parameters are the VEVs of the Wilson loops.

\subsubsection{Confined phase}

Let us first consider the phase in which all Wilson loops have a vanishing VEV, which corresponds to a confining phase. Generically, matter operators can have non-zero VEVs in this phase, breaking the 2d gauge symmetry to a $\mathbb{Z}_N$ subgroup. From the point of view of the IR theory, where the quiver defines a 2d $\mathbb{Z}_N$ lattice gauge theory, this leftover $\mathbb{Z}_N$ is the whole gauge group, as baryons are neutral under it. These VEVs trigger symmetry breaking of the 0-form global symmetry $G_0$, while the 1-form $\mathbb{Z}_N$ global symmetry remains intact. 

In 4d language, this phase corresponds to the baryonic branch. In this phase, for large $L_1$ and $L_2$, the quiver deconstructs the 4d non-local quantum theory known as $(1,1)$ little string theory, as explained in \cite{Arkani-Hamed:2001wsh}. The 2d 1-form $\mathbb{Z}_N$ global symmetry and the standard 4d 1-form $\mathbb{Z}_N$ global symmetry can be combined into a larger, 6d 1-form $\mathbb{Z}_N$ symmetry, which corresponds to that of the 6d theory \cite{Bhardwaj:2020phs}. 

Note that the case of large $L_1$ with $L_2=1$ is special in that the latticized space is a circle. Therefore, the quiver deconstructs the 5d maximally SUSY Yang-Mills theory, which in turn is UV completed by the 6d $(2,0)$ theory.\footnote{The instanton particles of the 5d maximally SUSY YM theory are the Kaluza-Klein tower of the compactification of the 6d $(2,0)$ theory on the M-theory circle.} In this case, the 1-form global symmetry is to be identified with the reduction of the $\mathbb{Z}_N$ 2-form global symmetry on the M-theory circle.

\subsubsection{Deconfined phase}

In the deconfined phase, Wilson loops take non-zero VEVs. In 4d language, this phase corresponds to the mesonic branch. From the 2d point of view, the non-zero VEVs of the Wilson loops result in the breaking of the 2d $\mathbb{Z}_N$ global 1-form symmetry. 
Schematically, assuming generic VEVs for the fields $A_p,\,B_p,\,C_p\sim \Lambda$, the VEV of a meson is given by $\langle \mathcal{M}\rangle \sim \Lambda^n$, with $n$ the number of constituent fields. From the 2d point of view, $n$ is proportional to the length $\ell$ of the Wilson loop. Therefore, the VEVs of Wilson loops satisfy a perimeter law as expected in a deconfined phase. 

\vspace{.5cm}

Both phases come together at the point where all VEVs vanish. From the 4d point of view, this is the origin of the moduli space, where the CFT lives. Thus, the 4d CFT appears as the confinement/deconfinement phase transition.

\section{Conclusions}

We have extended the proposal in \cite{Razamat:2021jkx} in various directions. First, we have explained that it is possible to associate a 2d $U(1)$ lattice gauge theory living on a torus to a 4d quiver gauge theory in which all gauge groups are $SU(N)$, with the lattice literally given by the quiver diagram. The 2d model has fractonic matter excitations \cite{Razamat:2021jkx} -- corresponding to baryons in 4d -- of charge $N$ under the 2d $U(1)$ gauge symmetry. Thus, in addition to the 0-form global symmetries discussed in \cite{Razamat:2021jkx}, the model has a $\mathbb{Z}_N$ global symmetry under which Wilson loops -- corresponding to mesons in 4d -- are charged. This is nicely encoded in the chiral ring relations, which allow $N$ Wilson lines to end on a charge $N$ matter field. Furthermore, the chiral ring relations show that these Wilson loops can move freely in the 2d world. 

Our construction falls into the general spirit of deconstruction. From this perspective, the deconstructing theory would be the 2d lattice gauge theory once the ``confining" dynamics of the $SU(N)$ nodes is taken into account. Interestingly, this point of view naturally fits with \cite{Pai:2019hor}, where it was argued that the confining dynamics of gauge fields is responsible for the fractonic behavior for confining strings. However, our quivers are actually conformal, rather than confining. It is the strong coupling dynamics of the $SU(N)$ gauge symmetry what ``confines" bifundamentals into baryons, which may be regarded as the confining strings that exhibit fractonic behavior. Deconstruction is explicitly realized, though, when moving onto the baryonic branch and introducing a scale associated to corresponding non-zero VEVs for bifundamentals. 

The 2d $\mathbb{Z}_N$ 1-form global symmetry characterizes the different phases of the lattice theory. In the deconfined phase/mesonic branch, Wilson loops exhibit perimeter law and break the 2d global 1-form symmetry. Alternatively, in the confined phase/baryonic branch, Wilson loops have vanishing VEVs and, correspondingly, the 2d global 1-form symmetry is unbroken. In this phase, the 2d and 4d $\mathbb{Z}_N$ 1-form global symmetries can be combined into the 6d $\mathbb{Z}_N$ 1-form global symmetry of a deconstructed 6d quantum field theory. 

It is interesting to consider this process from the point of view of the IR theory. From that perspective, one would simply say that most $U(1)$'s are anomalous. As discussed, from each of them only a $\mathbb{Z}_N$ remains non-anomalous and can be actually identified with actions of the center of the $SU(N)^{L_1\times L_2}$ gauge symmetry. From the quiver point of view, these transformations are local in the lattice (as well as in 4d), and thus give rise to a lattice $\mathbb{Z}_N$ gauge theory leading to a $\mathbb{Z}_N$ 1-form global symmetry in 2d. Thus, from this point of view, it is the center of the gauge group which, in the 4d or 2d guise, gives rise to the 6d $\mathbb{Z}_N$ 1-form global symmetry.

It is natural to expect that the ideas discussed in this work generalize to quivers on more general surfaces \cite{Franco:2012mm} or in higher dimensions \cite{Franco:2015tya,Franco:2021elb}. It would be interesting to investigate whether they lead to interesting condensed matter applications.

\section*{Acknowledgements}

D.R-G thanks G.Arias-Tamargo and O.Bergman for useful conversations. S.F. is supported by the U.S. National Science Foundation grants PHY-1820721, PHY-2112729 and DMS-1854179. D.R-G. is partly supported by Spanish national grant MINECO-16-FPA2015- 63667-P as well as the Principado de Asturias grant FC-GRUPIN-IDI/2018/000174.

\end{document}